# Step bunching and macrostep formation in 1D atomistic scale model of unstable vicinal crystal growth

F. Krzyżewski[1,*], M. Załuska-Kotur[1,2], A. Krasteva[3], H. Popova[4] and V. Tonchev[4]

[1]*Institute of Physics, Polish Academy of Sciences, al. Lotników 32/46, 02-668 Warszawa, Polska, Warsaw, Poland.*
[2]*Faculty of Mathematics and Natural Sciences, Card. S. Wyszynski University, ul. Dewajtis 5, 01-815 Warsaw, Poland*
[3]*Institute of Electronics, Bulgarian Academy of Sciences, 72 Tzarigradsko chaussee blvd, 1784 Sofia, Bulgaria*
[4]*R. Kaischew Institute of Physical Chemistry, Bulgarian Academy of Sciences, Acad. G. Bonchev Str., block 11, 1113 Sofia, Bulgaria*

[*]Corresponding author: fkrzy@ifpan.edu.pl

**Abstract**. We devise a new 1D atomistic scale model of vicinal growth based on Cellular Automaton. In it the step motion is realized by executing the automaton rule prescribing how adatoms incorporate into the vicinal crystal. Time increases after each rule execution and then $n_{DS}$ diffusional updates of the adatoms are performed. The increase of $n_{DS}$ switches between the diffusion-limited (DL, $n_{DS}=1$) and kinetics-limited (KL, $n_{DS} \gg 1$) regimes of growth. We study the unstable step motion by employing two alternative sources of instability – biased diffusion and infinite inverse Ehrlich-Schwoebel barrier (iiSE). The resulting step bunches consist of steps but also of macrosteps since there is no step-step repulsion incorporated explicitly into the model. This complex pattern formation is quantified by studying the time evolution of the bunch size $N$ and macrostep size $N_m$ in order to find the proper parameter combinations that rescale the time and thus to obtain the full time-scaling relations including the pre-factors. For the case of biased diffusion the time-scaling exponent $\beta$ of $N$ is 1/2 while for the case of iiSE it is 1/3. In both cases the time-scaling exponent $\beta_m$ of $N_m$ is ~$3\beta/4$ in the DL regime and $3\beta/5$ in the KL one.

## Introduction

The production of new devices nowadays reaches new frontiers of miniaturization but these fast and dramatic changes require very precise tuning of layer-by-layer (step-flow) crystal growth. Thus the detailed use and directed manipulation of the processes and patterns on atomic scale is of crucial practical relevance. For this it is necessary to reach a fundamental understanding of the growth mechanisms and their consequences on atomic scale. This is why surface morphologies resulting from various kinds of crystal growth processes are subject of interest for large groups of researcher [1-5]. Surface self-organization resulting in well-ordered structures is used to build templates for growing nano-scale objects such as nano-dots or nano-wires [6, 7]. It is known that at the miscut surfaces the asymmetry between adatom fluxes which attach to the steps from above and bottom terraces leads to surface instabilities [8-12]. If the amount of particles attaching to the step from the lower terrace is higher than particle flux from the above terrace meandered patterns emerge [10]. Otherwise, when flux incoming from above is higher step bunching process happens [10, 13]. Flux asymmetry at steps can be induced by various dynamic mechanisms. The most often discussed in this context are biased adatom diffusion i.e. due to electromigration [10, 14] or the existence of Ehrlich-Schwoebel barrier (SB) [1, 3-5, 15]. Below we will concentrate on these two sources of the surface instability.

The mechanism of step bunching that happens due to each of these instability sources has been widely discussed and analyzed in its various aspects [1, 3-5, 13-18]. Its initial stages, starting from step doubling are easy to observe and analyze. However, when it comes to exact evaluation of scaling factors it is necessary to ensure large systems, many samples and long times of study. It is very difficult in experimental systems [14] and in more realistic MC simulation as well [5, 10]. Analytic models give better chance [10, 17-22], however it would be good to link their parameters with the ones of discrete systems. From the other side classification of the studied phenomena to the proper universality class[19] is a good course to understand mechanisms and character of this dynamical process. In this work a model based on cellular automata (CA) is proposed as a simple, clear and powerful tool that is expected to be able to go beyond the analytical treatments. By extensive investigations of the proposed model in one dimensional system we are able to determine different scaling of the bunch size $N$ with time in the regime of intermediate asymptotics [23]. In the case of biased diffusion we get scaling exponent $\beta=1/2$ and in the case of the surface with infinite inverse Ehrlich-Schwoebel barrier (iiSB) it is $\beta=1/3$. Step bunches we observe in our simulations consist of single steps that have size of one unit cell, but also of macrosteps with size of multiple unit cells. Such formations are seen in experiments [24] and their time evolution remains subject of studies [25, 26]. Macrosteps are created during the surface evolution process because there is no step-step repulsion incorporated into the model. This complex pattern formation is analyzed and together with the scaling of bunch size $N$ the time dependence of the size of macrosteps $N_m$ is studied. In both variants of the model the time scaling exponent $\beta_m$ of $N_m$ is reduced o approximately $3\beta/4$ of the scaling exponent of $N$ in the DL regime and $3\beta/5$ in the KL one. We find also the proper parameter combinations that rescale the time and as an effect all studied curves are collected along an universal one.

Both studied systems allow to investigate scaling in wide range of parameters. Preliminary results presented in [27] shown some examples of the scaling behavior. Now we expand our analysis into other areas of parameters and show how results for wide range of model parameters scale along universal curves. We checked that different bias values give the same scaling. More interestingly system with iiSB the data for he bunch size also scales along the same line in both slow and fast diffusion limit, despite the fact that the first one means kinetics limited and the second diffusion limited growth.

Below the model is described, then we show and analyze time scaling for step bunch sizes and macrostep sizes for both studied systems. General scaling functions are shown and discussed. We also compare step profiles forms as an effect of biased diffusion and the presence of iiSB at the step.

**The model**

The model is devised as the simplest possible proposition that is able to achieve atomic scale resolution still retaining the possibility for fast calculations on large systems. It is built as a combination of two modules: the deterministic one – a cellular automata (CA) and a diffusional module - a typical Monte Carlo (MC) procedure that brings the concept of stochasticity into the model. Each cell from the 1D colony is given a value equal to its height in the vicinal stairway that descends from the left to the right. In the beginning, the steps are regularly distributed at distance $l_0$. Another 1D array of the same size $L$ contains the adatoms. In the beginning they are randomly distributed over the surface with concentration $c$. The growth rule defines that each time there is an adatom at the right nearest neighboring site to a step or macrostep it attaches unconditionally to it. Then the step or the lowest layer of the macrostep advances one position to the right what is realized by increasing the value of the vicinal cell colony at the position of the adatom by one as illustrated in Fig. 1. The adatom is deleted from the adatoms array. The growth updates using the growth rule are performed in a parallel fashion – the update (change of height at that position with 0 or 1) of each cell from the vicinal crystal is kept aside in a mirror array while every cell is being checked, then the whole cell population is renewed at instance using this mirror array and only then the time is increased by 1. Each execution of the automaton rule is complemented by compensation of the adatom concentration to $c$ and the adatom population is then subjected to diffusional update(s) in a serial manner, their number is denoted by $n_{DS}$. In any diffusional update a total number of adatoms positions equal to the size of the adatom array $L$ is chosen sequentially at random. Then, if adatom is found there, it is tried to jump left or right with some probabilities, usually their sum being 1, except in the case of iiSE, the move is accepted only if the next chosen adatom position is not occupied already by an adatom. The change of this adatom's position is enforced without postponing and another position is checked for availability of an adatom. These diffusional updates do not contribute to the increase the time. Thus, with increasing $n_{DS}$ is realized a transition from diffusion-limited (DL) growth to a kinetics-limited (KL) one – while the kinetic events (growth rule executions) happen with the same fixed frequency the diffusing adatoms can make on average as many hops as determined by $n_{DS}$ before being eventually captured by the growing surface[28]. The diffusion is influenced by one of two principal sources of instability – directional bias or iiSE. The bias is realized by defining that the hop probability to the right as (0.5 + δ) while to the left it is (0.5 - δ). The iiSE is realized through inhibition of the diffusional hops to the left when the adatom is right next nearest neighbor to a step or macrostep and inhibition of the diffusional hops to the right when the adatom is right nearest neighbor to a step or macrostep. The growth rule for iiSE case is presented in Fig 1b. Adatom diffusion over barrier outlined there is blocked. Note the iiSE turns the model into one-sided. Destabilizing factors in the model are not opposed by step-step repulsions, hence there is no factor preventing formation of macrosteps. Indeed as it will be shown later step bunches in fact are build out of macrosteps that become dominating structure visible in the profile of the crystal.

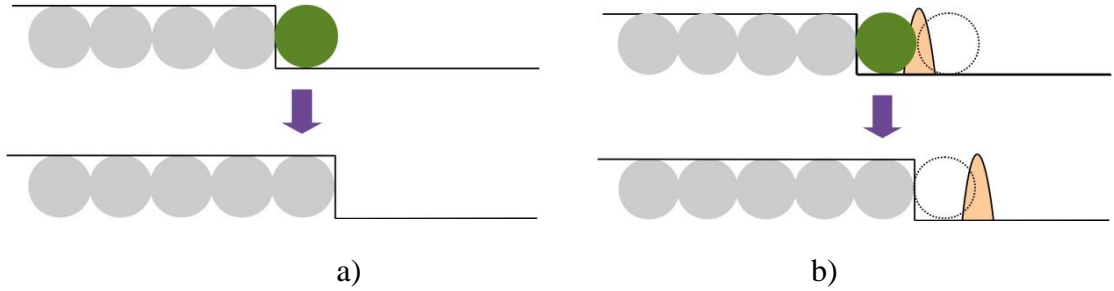

a)                                      b)

Fig 1. Step move as an effect of CA module incorporation of diffusing particle into the step a) in the biased diffusion model and b) in the system with iiSE. It can be seen how the infinite ES barrier is removed from the place where it is established and re-established at one lattice position to the right.

The model permits fast calculations on systems with large sizes thus achieving the regime of intermediate asymptotics [23] where a reliable statistics is collected for the monitored properties. In order to control the developing surface patterns we adopt a modification [27] of an established monitoring protocol [29]. It is the formation of macrosteps that determines the need of this modification. We investigate time dependent bunch size $N$ and macrostep size $N_m$, which are the parameters useful for description of step bunching phenomenon. Important criterion build into the protocol defines when two neighboring steps belong to the same bunch – it is when the distance between them is less than $l_0$, whereas groups of steps with distance $l=0$ are considered as macrosteps. Bunch size $N$ measures the height interval between the topmost bunch step and the lowest. The same approach is applied to the macrostep size $N_m$. In numerical results presented below bunch usually consists of steps and macrosteps as well (see Fig. 2). To obtain proper scaling in the results presented below we performed calculations for large systems (up to 180 000 sites), large number of time steps (~ $10^8$) and usually repeat calculations at least 5 times each.

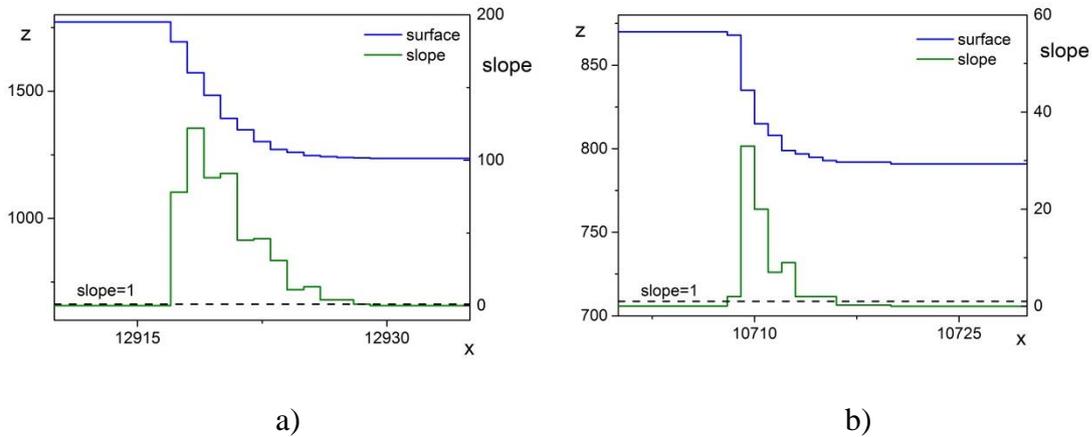

a)                                      b)

Fig. 2 Bunch shape for a) biased case with $\delta=0.05$ and b) iiSB model evolved in $10^7$ time steps and for $c_0=0.2$, in both cases $n_{DS} = 1$. Top curves show surface height and bottom curves denote local surface slope.

## Results and discussion

The profiles that build up at the vicinal crystal surface after some time are shown for two studied instability types in Fig. 2a and b. As can be seen both types of instabilities lead to the creation of bunches that are built out of macrosteps and mono-steps. It can be more precisely seen in the plots for the surface slopes below. Slopes are calculated from the discrete data as $s=h/l$, where $h$ is the height of macrostep and consequently $h=1$ for each single step and $l$ is the length of the terrace at the right step side. Such way of the surface slope presentation allows to illustrate overall bunch shape, in particular to find where the part of the highest slope is present. In these plots dashed curves separate closely located macrosteps (slopes above 1) and steps (slopes below 1). In both cases single steps are present only at the bunch edges and there are only few of them. It seems that they provide bunch communication by detaching at one bunch and attaching to the another one. The profiles of the step bunched surfaces are identical to the ones obtained in [29] – the bunches are steepest in one of the ends where the steps join the bunch from behind. This type of step bunching requires extension of the step bunching classification [30].

Such interpretation of the shape of surface profile can be supported by the analysis of the trajectories of steps and macrosteps plotted in Fig. 3. In this plot macrosteps trajectories are presented by thick, red lines and single step trajectories by dotted blue lines. We can see that most steps are collected into macrosteps. Moreover each thick line in Fig. 3 in fact consists of several macrostep lines that are so close together, building one bunch that they are seen as one trajectory. During the process of crystal growth single steps detach from one of macrostep and moving much faster reach and attach to the preceding macrostep. Within the same period of time evolution such events are more frequent in the case of iiSB system than in the biased one. It is also clear that macrosteps in this last case are larger, because there are only three of them in the whole system at $t=5 \cdot 10^6$, whereas at the same time for iiSB system there are seven separated macrosteps.

Bunch shapes in both cases as seen in Fig 2. are similar. The main difference is that macrosteps and as an effect bunches in biased system are much higher (see the slope scale at the right hand side), but less steep. The maximal slope is moved left in both cases, what means that the bunches are steeper at the beginning. In the case of iiSB system this slope maximum is almost at the beginning of the bunch whereas in biased system the maximal slope seems to be distributed over larger distance and moved slightly from the beginning.

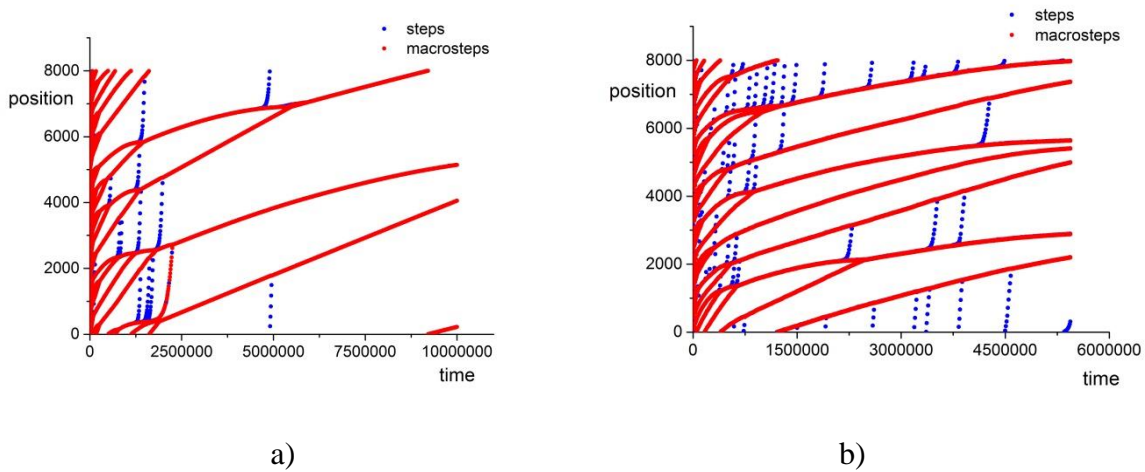

a)          b)

Fig 3. Step and macrostep trajectories for a) biased model with $\delta = 0.1$ and b) iiSE model with $c=0.2$

We want to characterize the character of the surface shape transformations during crystal growth process. We concentrate on of the time dependence of the mean bunch size. Let us first analyze biased system. We studied systems for different values diffusional updated $n_{DS}$ and bias $\delta$. Both parameters characterize the type of surface dynamics. We also studied crystal growth for different initial vicinal distance $l_0$ and for changed adatom concentrations $c_0$. The first parameter corresponds to the surface miscut while $c_0$ is related to the supersaturation at the step. It appears that all curves for the mean bunch size $N$ and values of $n_{DS}$ not bigger than 30 can be plotted along one, master curve when time $t$ is recalled as follows

$$T = \frac{\Omega \delta c_0 n_{DS}}{l_0} t \tag{1}$$

where $\Omega$ is the constant area of elementary cell. In this representation time $T$ is dimensionless. The master curve after time rescaling is shown in Fig. 4a. All data calculated for a wide extent of simulation parameters lie along single straight line when plotted in a log-log graph. The slope of line $\beta=1/2$, what means that bunch size scales with this exponent for all values of bias and independently from the other parameters.

In Fig. 4b time dependence of the mean macrostep size $N_m$ is presented. The time scaling exponent $\beta_m \approx 3/8$ for the DL regime can be expressed as $\beta_m \approx 3\beta/4$ while for the KL-regime $\beta_m = 3\beta/5$.

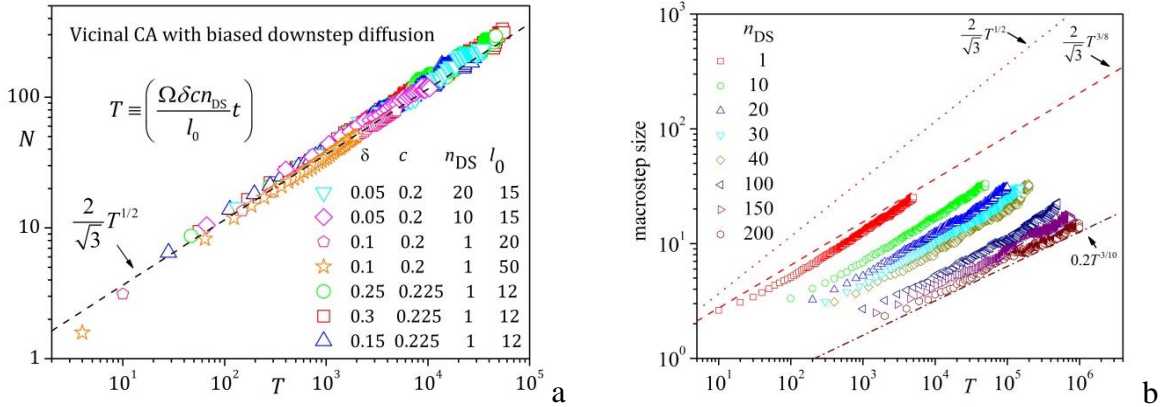

Fig 4. Time scaling of a) bunch size $N$ and b) size of macrostep $N_m$ for different parameters in the case of biased system.

When, instead of external bias, infinite inverse Ehrlich-Schwoebel barrier is present the time scaling of $N$ is different. All curves obtained for different growth parameters again can be rescaled into one master curve. We can see this in Figure 5a. In the case of iiSB the universal time is

$$T = \frac{\Omega c_0 n_{DS}}{l_0^2} t. \tag{2}$$

The slope of this curve is lower than the slope obtained in biased system and gives time-scaling exponent of $N$ $\beta = 1/3$. All previous observations about slower process of bunch formation in this system summarize in this value of the scaling exponent. Moreover we can see that the $l_0$ dependence of the scaling relation is different in both studied cases. When we rescale time dependence of multistep size $N_m$ for the iiSB system using (2) again curves for different number of diffusion steps separate as presented. Thus, in the DL regime $\beta_m=1/4$ while in the KL one $\beta_m=1/5$. It reproduces the same as above relations between exponents -

$\beta_m \approx 3\beta/4$ )DL) and $\beta_m = 3\beta/5$ (KL). Difference in scaling exponents for $N$ and $N_m$ is an effect of internal bunch structure. Bunches consist of mono- and macrosteps. Size of bunches increases faster because this dynamics includes two different processes of macrostep growing and of drawing them together.

We can see that both bunching mechanisms, studied within one picture express their different nature. Even if it is easy to represent both of them by the inequivalent particle fluxes they do not become equal. What is worth noting is that time-scaling exponent $\beta$ of the bunch size $N$ does not depend on the number of diffusional steps in all studied cases. However evolution of bunches and macrosteps scale differently. Above we analyze only infinite inverse Ehrlich-Schwoebel barrier because finite barriers lead to other, new effects. The work about these effects is in progress.

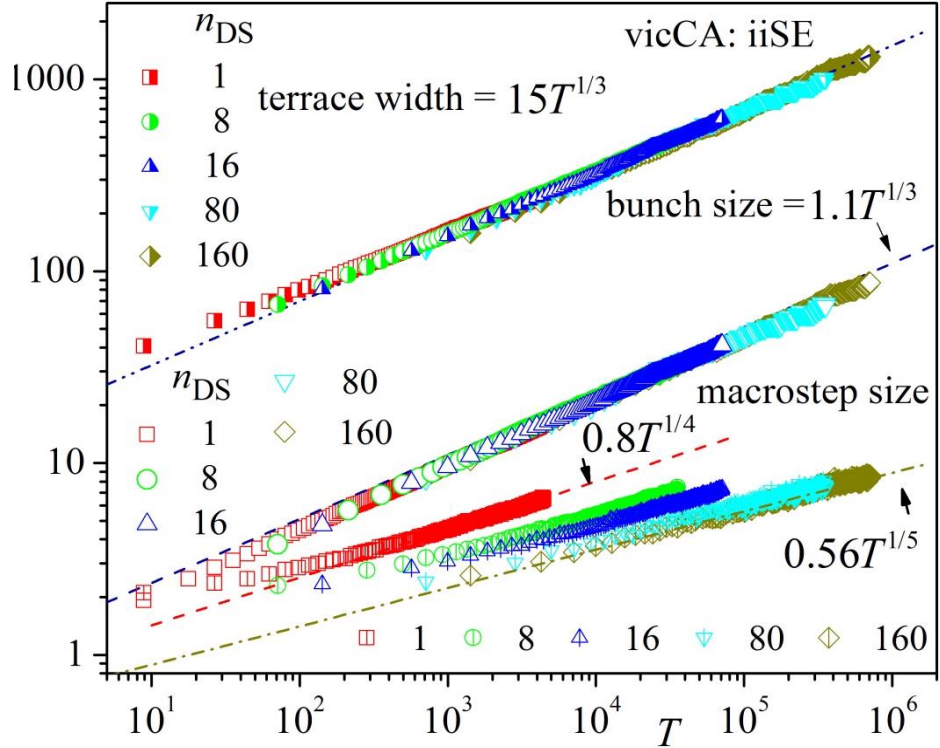

Fig 5. Time scaling of terrace size, bunch size $N$ size of macrostep $N_m$ for different parameters in the case iiSE.

**Conclusions**

In order to explore the process of surface patterning in unstable vicinal growth we build a new model based on a combination of Cellular Automata whose rule prescribes strictly the growth events and a random part modelling the diffusion of the adatoms. Our model generates bunches that consist of mono- and macrosteps but treats them on the same ground what is not valid for the other models we know. We believe that our model bares the essence of the crystal growth being combination of ordering processes complemented by disordering ones. We use the model to study two instabilities with alternative sources - biased diffusion and iiSE in the regime of intermediate asymptotics [23] where the surfaces are self-similar (self-affine) both in space and in time. In order to quantify this self-similarity we obtain time-scaling relations for the bunch size *N* and size of macrosteps. The whole range of parameters is covered, this permits to find the proper parameter combinations to rescale the time and to decipher the time-scaling relations down to the numerical pre-factor. Interestingly two studied sources of the bunch instability - bias and iiSE manifest entirely different time evolution. Moreover the time scaling they show is completely insensitive to parameter changes. All biased systems with $n_{DS}<40$ build bunches as $1.1T^{1/2}$ regardless of the bias and number of diffusion steps we apply. For values of $n_{DS}$ bigger than 40 the time-scaling exponent of 1/2 is preserved but another rescaling of the time should be applied in order to permit collaps of all date on the same curve. Differently, bunch sizes for all studied values of $n_{DS}$ up to 200 scale as $1.1T^{1/3}$. What distinguishes in between the two regimes of growth is the time-scaling exponent of the macrostep size. Such universal behavior has to have simple explanation expressed in some general rules that apply in the studied systems. Rules should differ between two presented mechanisms of instability in some obvious way. We believe that such relations can be found soon and that they are also true for real crystal surfaces. Our model is easily generalizable to higher dimensions where new phenomena are expected – bunching of straight steps, step meandering of equidistant steps and simultaneous step bunching and meandering[4, 31, 32].

**Acknowledgements**

This work is supported by the Bulgarian NSF, grant No. T02-8/121214, and a bilateral project between Bulgarian and Polish Academies of Sciences. F.K. acknowledges NCN of Poland, grant No.2013/11/D/ST3/02700, and a fellowship under Erasmus+. VT is grateful for the very stimulating working conditions in the Institute of Physics of the Polish Academy of Sciences.